\documentclass{PoS}

\usepackage{epsfig,multicol}
\usepackage{amssymb,amsmath,bm, mathrsfs}
 \usepackage{amsfonts}
\usepackage{amstext}
\usepackage{epsf}
\usepackage{graphicx}
\usepackage{longtable}
\usepackage{afterpage}
\usepackage{placeins}
\usepackage{color}
\usepackage{bbm}
\usepackage{afterpage}

\usepackage{slashed}
\usepackage{textcomp}
\usepackage{placeins}

\newcommand{\gev}{\, {\rm GeV}}

\allowdisplaybreaks[2]
\newcommand{\be}{\begin{equation}}
\newcommand{\ee}{\end{equation}}
\newcommand{\bea}{\begin{eqnarray}}
\newcommand{\eea}{\end{eqnarray}}
\newcommand{\nn}{\nonumber}
\newcommand{\bi}{\begin{itemize}}
\newcommand{\ei}{\end{itemize}}

\definecolor{orange}{rgb}{1,0.4,0.1}

\DeclareMathOperator{\diag}{diag}

\newcommand{\lsim}{
\mathrel{\hbox{\rlap{\hbox{\lower4pt\hbox{$\sim$}}}\hbox{$<$}}}}
\newcommand{\gsim}{
\mathrel{\hbox{\rlap{\hbox{\lower4pt\hbox{$\sim$}}}\hbox{$>$}}}}

\title{Flavoured Dark Matter Beyond MFV}

\ShortTitle{Flavoured Dark Matter Beyond MFV}

\author{\speaker{Monika Blanke}\vspace{1mm}\\
        CERN Theory Division, CH-1211 Geneva 23, Switzerland\vspace{1mm}\\
Institut f\"ur Theoretische Teilchenphysik, Karlsruhe Institute of Technology,
Engesserstra{\ss}e 7,\\ D-76128 Karlsruhe, Germany\vspace{1mm}\\
Institut f\"ur Kernphysik, 
Karlsruhe Institute of Technology,
Hermann-von-Helmholtz-Platz 1,\\ 76344 Eggenstein-Leopoldshafen, Germany\vspace{1mm}\\
        E-mail: \email{monika.blanke@kit.edu}}

\abstract{We review a model of quark flavoured dark matter with new flavour violating interactions. This simplified model describes Dirac fermionic dark matter that is charged under a new $U(3)$ flavour symmetry and couples to right-handed down quarks via a scalar mediator. The corresponding coupling matrix is assumed to be the only new source of flavour violation, which we refer to as the Dark Minimal Flavour Violation (DMFV) hypothesis. This ansatz ensures the stability of dark matter. We discuss the phenomenology of the simplest DMFV model in flavour violating observables, LHC searches, and direct dark matter detection experiments. Especially interesting is the non-trivial interplay between the constraints from the different sectors.}

\FullConference{Flavorful Ways to New Physics - FWNP,\\
		28-31 October 2014\\
		Freudenstadt - Lauterbad, Germany}

\begin{document}

\section{Introduction}

The existence of dark matter (DM) is one of the very few solid indications that physics beyond the Standard Model (SM) must exist. Various astrophysical and cosmological observations tell us that the visible baryonic matter constitutes only a small fraction of the total energy density in the universe. Roughly five times as much DM exists which until now has manifested itself only through its gravitational interactions. 
At the same time very little is known on the particle nature of DM. We don't know through what interactions DM couples to the SM, we have no experimental information on its spin, or its mass. It appears too appealing to be a coincidence though that a weakly interacting massive particle (WIMP), i.\,e.\ a weak-scale particle with weak interactions naturally leads to the right relic density of $\Omega_\text{DM}h^2 \simeq 0.12$, a theoretical observation often referred to as the WIMP miracle.

Adopting the WIMP assumption in what follows, we are still left with a lot of freedom on the structure of the dark sector. For instance similarly to the visible sector, DM could come in multiple generations which are distinguished through its mass. This idea of flavoured DM has become quite popular in recent years (see \cite{Kile:2013ola} for a review). Yet most studies so far assumed that the DM sector satisfies the Minimal Flavour Violation (MFV) assumption \cite{Buras:2000dm,D'Ambrosio:2002ex}, i.\,e.\ that no new sources of flavour violation are present beyond the SM Yukawa couplings. While this assumption saves well-measured flavour changing neutral current (FCNC) observables from dangerously large new contributions, at the same time it kills the possibility to detect flavoured DM in rare flavour violating decays, and it prevents us from explaining various small tensions in the present flavour data.

\begin{figure}[h!]
\centering
\includegraphics[width=0.35\textwidth]{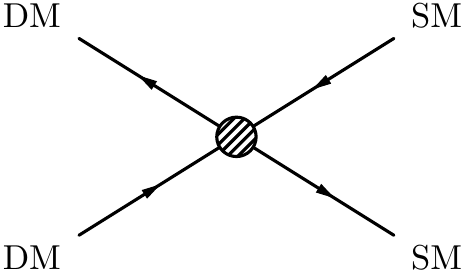}\qquad\qquad
\includegraphics[width=0.35\textwidth]{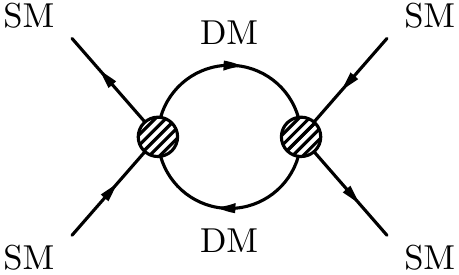}
\caption{Schematic diagrams contributing to experimental constraints on flavored DM.}
\label{fig:DMloop}
\end{figure}

In \cite{Agrawal:2014aoa} we took a different approach. We constructed a simplified model of flavoured DM, where the coupling to SM quarks explicitly constitutes a new source of flavour and CP violation. In addition to the usual DM discovery channels in direct and indirect detection experiments and collider searches, c.\,f.\ left diagram in figure \ref{fig:DMloop}, sizeable contributions to FCNC observables are then also expected, as depicted in the right diagram in figure \ref{fig:DMloop}. The experimental constraints from meson-antimeson mixing observables can then be used to constrain the structure of the coupling between SM and DM sectors. In this article we review the model introduced in \cite{Agrawal:2014aoa}, discussing in turn its coupling and flavour structure and its phenomenology in flavour, DM and LHC experiments.

\section{Dark Minimal Flavour Violation}

We introduce DM as a Dirac fermion $\chi$ which is gauge singlet and  transforms under the triplet representation of a new global flavour symmetry $U(3)_\chi$. Its coupling to the SM quark sector, more explicitly to the right-handed down type quarks, is mediated by a scalar field $\phi$. While $\phi$ is singlet under $SU(2)_L$, it carries colour and hypercharge. The Lagrangian of this simplified model reads
\begin{eqnarray}
  {\cal L}&=& \mathcal{L}_\text{SM}+
  i \bar \chi \slashed{\partial} \chi
  - m_{\chi}  \bar \chi \chi  - (\lambda_{ij} \bar {d_{R}}_i \chi_j \phi + {\rm h.c.}) \nn \\
  && \qquad\,
  +
  (D_{\mu} \phi)^{\dagger} (D^{\mu} \phi) - m_{\phi}^2 \phi^{\dagger} \phi
  +\lambda_{H \phi}\, \phi^{\dagger} \phi\, H^{\dagger} H
  +\lambda_{\phi\phi}\, \phi^{\dagger} \phi\, \phi^{\dagger} \phi \,,\label{eq:Lagrangian}
\end{eqnarray}
where $i,j=1,2,3$ are flavour indices.

In order to keep the model minimal, we assume that the new Yukawa coupling $\lambda$ is the only new source of flavour symmetry breaking, in addition to the SM Yukawa couplings $Y_{u,d}$. Conceptually this ansatz extends the MFV principle to the dark sector flavour group, and we therefore call it {\it Dark Minimal Flavour Violation (DMFV)}. We stress that despite its conceptual analogy to the MFV hypothesis, the phenomenology is very different due to the new flavour violating coupling $\lambda$ whose structure is unrelated to the SM Yukawas. Consequently DMFV can lead to large new physics contributions to FCNC observables.

The DMFV hypothesis has a number of interesting implications. First, by construction the DM mass term $m_\chi$ in \eqref{eq:Lagrangian} is flavour universal. Higher order contributions in the DMFV expansion however generate a small mass splitting,
\be\label{eq:mchi}
  m_{\chi_i}
  =
  m_{\chi} (\mathbbm{1}  +\eta\, \lambda^\dagger \lambda+\dots)_{ii}\,.
\ee
Note that the DM mass matrix is aligned with the coupling matrix $\lambda$.

Second, the DMFV hypothesis ensures DM stability, in complete analogy to the stability of DM in  the MFV case \cite{Batell:2011tc}. The flavor
symmetry $U(3)_Q \times U(3)_u \times U(3)_d \times U(3)_\chi$
broken only by the Yukawa couplings
$Y_u$, $Y_d$ and $\lambda$, together with $SU(3)_\text{QCD}$ imply an
unbroken $\mathbbm{Z}_3$ symmetry, under which only the new particles
$\chi_i$ and $\phi$ are charged, and transform as
\be
  \chi_i \to e^{2\pi i/3} \chi_i, \qquad
  \phi \to e^{-2\pi i/3} \phi
  \ .
\ee
This symmetry prevents the decay of any of
these states into SM particles only, and therefore renders the
lightest state stable. 

Third, the DMFV assumption significantly reduces the number of free parameters. The $U(3)_\chi$ symmetry can be used to rotate away some parameters from the coupling matrix $\lambda$, and a convenient parameterisation is then given by
\be
\lambda = U_\lambda D_\lambda\,,
\ee
where $U_\lambda$ is a unitary matrix containing three mixing angles $\theta_{ij}^\lambda$ and three complex phases $\delta^\lambda_{ij}$ ($ij=12,13,23$) parameterised as in \cite{Blanke:2006xr}. $D_\lambda$ is a real and diagonal matrix, for which we use the parameterisations
\be\label{eq:Dlambda}
D_\lambda \equiv \diag(D_{\lambda,11},D_{\lambda,22},D_{\lambda,33})=
\lambda_0\cdot \mathbbm{1}+\diag(\lambda_1,\lambda_2,-(\lambda_1+\lambda_2))\,.
\ee

Before moving on we note that the model described by the Lagrangian \eqref{eq:Lagrangian} is the minimal mocel realising the DMFV assumption. We hence refer to it as the {\it minimal DMFV (mDMFV) model}. While for the subsequent phenomenological discussion we restrict our attention to the mDMFV model for simplicity, we note that many conclusions hold more generally within the DMFV framework.

\section{Flavour constraints}

\begin{figure}[h!]
\centering
\includegraphics[width=0.33\textwidth]{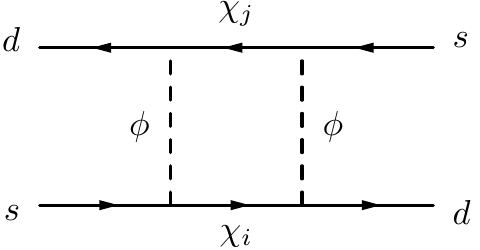}
\caption{New contribution to $K^0-\bar K^0$ mixing in the mDMFV model.}
\label{fig:kkbar}
\end{figure}

New flavour violating interactions at the weak scale are strongly constrained by $\Delta F = 2$ data, i.\,e.\ observables related to neutral meson mixing. In the mDMFV model new contributions to $K^0-\bar K^0$ mixing are generated by the box diagram in figure \ref{fig:kkbar}. Analogouse diagrams contribute to $B_{d,s}-\bar B_{d,s}$ mixing.

Exact expressions for the mDMFV contributions to $\Delta F =2$ observables can be found in  \cite{Agrawal:2014aoa}. Schematically it can be written as
\be
M_{12}^{K,\text{new}} \sim
  \left((\lambda\lambda^\dagger)_{sd}\right)^2 { F(x) }\,,\qquad 
M_{12}^{d,\text{new}} \sim
  \left((\lambda\lambda^\dagger)_{bd}\right)^2 { F(x) }\,,\qquad 
M_{12}^{s,\text{new}} \sim
  \left((\lambda\lambda^\dagger)_{bs}\right)^2 { F(x) }\,,
\ee
where $F(x)$ is the relevant loop function that is flavour independent. We see that the size of new contributions to FCNCs is determined by the off-diagonal elements of $(\lambda\lambda^\dagger)$. Consequently in order not to spoil the good agreement with the data, $(\lambda\lambda^\dagger)$ has to be close to diagonal.

\begin{figure}[h!]
\centering
\includegraphics[width=.5\textwidth]{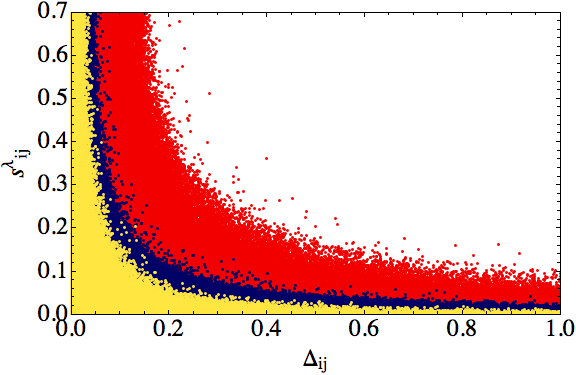}
\caption{\label{fig:deg}Allowed ranges for the mixing angles $s^\lambda_{ij}=\sin\theta^\lambda_{ij}$ as a function of the deviation from the $ij$-degeneracy line $\Delta_{ij}=|D_{\lambda,ii}-D_{\lambda,jj}|$. $ij=12$ in yellow, $ij=13$ in blue, $ij=23$ in red.}
\end{figure}

Such a structure can be achieved in various ways, as discussed in detail and checked numerically in \cite{Agrawal:2014aoa}. In short, if a sizeable mixing angle $\theta^\lambda_{ij}$ is present, then the entries $D_{\lambda,ii}$ and $D_{\lambda,jj}$ of the diagonal component of $\lambda$
have to be quasi degenerate. Consequently we end up with the five scenarios for the structure of $\lambda$ pointed out in \cite{Agrawal:2014aoa}:
\begin{enumerate}
\item Universality scenario -- The diagonal coupling matrix $D_\lambda$ is nearly universal and large mixing angles in $U_\lambda$ are allowed.
\item Small mixing scenario  -- The mixing matrix $U_\lambda$ is close to the unit matrix, while $D_\lambda$ is arbitrary.
\item $ij$-degeneracy scenarios ($ij=12,13,23$) -- Two elements of $D_\lambda$ are almost equal, $D_{\lambda,ii}\simeq D_{\lambda,jj}$, and only the corresponding mixing angle $s^\lambda_{ij}$ is allowed to be large.
\end{enumerate}
Due to experimental and theoretical uncertainties, some deviation from exactly degenerate couplings or vanishing mixing angles is of course possible. Figure \ref{fig:deg} quantifies the allowed amount for the three cases $ij=12,13,23$. We see that the $23$ system is least constrained, due to the weaker bounds coming from $B_s-\bar B_s$ mixing.

The new contributions to rare $K$ and $B$ decays on the other hand are small. While this is welcome in the case of the very constraining decays $B\to X_s\gamma$ and $B_s\to\mu^+\mu^-$, the mDMFV model can not explain the tension observed in $B\to K^*\mu^+\mu^-$ and does not predict any interesting effect in very clean decays like $K\to\pi\nu\bar\nu$. 
Similarly also the non-standard contributions to electroweak precision observables and electric dipole moments turn out to be small.

\section{Dark matter phenomenology}

While the flavour phenomenology gives us a lot of information on the structure of the coupling matrix $\lambda$, it cannot constrain the mass spectrum of the three dark generations, $m_{\chi_i}$. In order to proceed with the DM phenomenology we therefore have to make a choice which of the three states is the lightest and constitutes the DM. We restrict our attention to the case of $b$-flavoured DM, i.\,e.\ the lightest state couples dominantly to the $b$ quark for the following reasons. The constraints from direct detection experiments become much less stringent than what would be the case for a strong couplings to $d$ quarks. Similarly also the bounds from LHC searches are easier to fulfill. Additional benefits of $b$ flavoured DM are the potential for interesting $b$-jet signatures at the LHC and the possibility to explain the excess $\gamma$-rays observed at the galactic center \cite{Agrawal:2014una}.

Further we require the DM to be a thermal relic. This assumptions sets constraints on the DM coupling as a function of the DM and the mediator mass. Depending on the mass splitting and the resulting life time of the heavier states
this affects either only the coupling $D_{\lambda,33}$ or also the couplings of the first two generations, if they are quasi-degenerate with the DM. The corresponding constraints are imposed on the parameter space in the subsequent analysis.

\begin{figure}[h!]
  \includegraphics[width=0.3\textwidth]{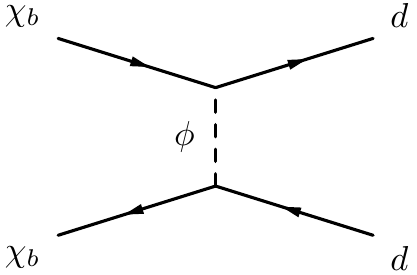}\hfill
  \includegraphics[width=0.3\textwidth]{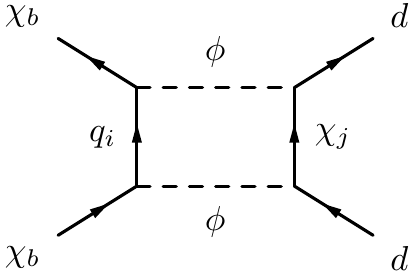}\hfill
  \includegraphics[width=0.3\textwidth]{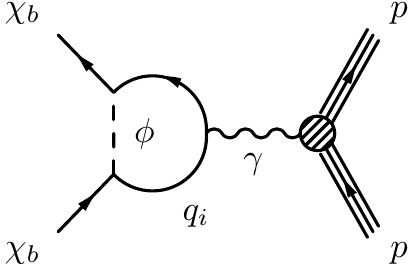}
  \caption{Diagrams contributing to WIMP-nucleon scattering in the mDMFV model.}
  \label{fig:dd}
\end{figure}

Several diagrams contribute to direct DM detection, as shown in figure \ref{fig:dd}. Firstly it is interesting to note that with the precision achieved by present experiments, mainly LUX \cite{Akerib:2013tjd}, we have become sensitive to DM quark scattering at the one loop level. Therefore even if the tree level diagram is suppressed by a small mixing angle $\theta^\lambda_{13}$, large  contributions to direct detection experiments are still present that need to be controlled.

Furthermore the one  loop box and photon penguin diagrams shown in figure \ref{fig:dd} can not be suppressed by small flavour mixing angles. Luckily however it is their intrinsic structure that supplies us with a suppression mechanism. The photon penguin receives an enhancement factor $\log\left(m_{q_i}^2/m_\phi^2\right)$ and therefore comes with a relative minus sign with respect to the box contribution. It can be shown (see \cite{Agrawal:2014aoa}) that the destructive interference becomes effective if the coupling $D_{\lambda,11}$ lies in a certain range.

\begin{figure}[h!]
\includegraphics[width=.49\textwidth]{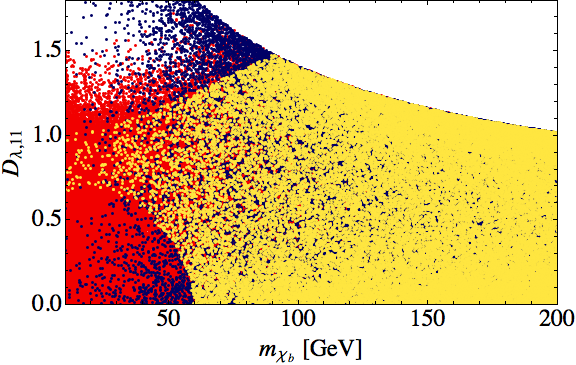}\hfill
\includegraphics[width=.49\textwidth]{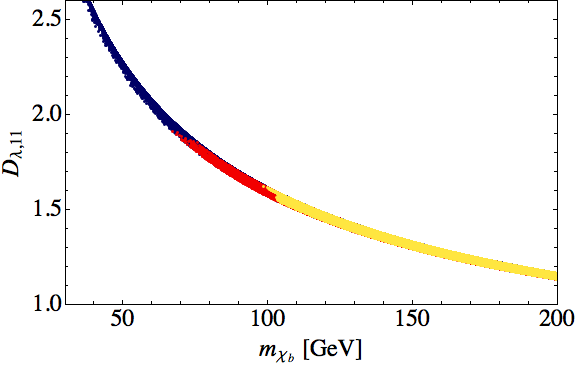}
\caption{\label{fig:mchi-D11}First generation coupling $D_{\lambda,11}$ as functions of the DM mass $m_{\chi_b}$ for significant mass splitting $m_{\chi_{d,s}}>1.1m_{\chi_b}$ scenario (left), and in the $13$-degeneracy scenario $m_{\chi_{d}} \simeq m_{\chi_b}$ (right). The red points satisfy the bound from LUX, while the blue points satisfy the $\Delta F=2$ constraints. For the yellow points both LUX and $\Delta F =2$ constraints are imposed.}
\end{figure}

This effect can be observed in figure \ref{fig:mchi-D11}. While for large DM masses $m_{\chi_b}\gsim 100\gev$ only the thermal relic constraint restricts the allowed range for $D_{\lambda,11}$, for smaller DM masses the LUX bound becomes effective and restricts $D_{\lambda,11}$ to lie in a certain range. Interestingly this constraint is only apparent if DM and flavour constraints are taken into account simultaneously. The reason is that the destructive interference necessary to comply with LUX data can in principle also be obtained for different $D_{\lambda,11}$ values; this however requires both a coupling non-degeneracy $D_{\lambda,11}\ne D_{\lambda,22}$ and a large mixing angle $\theta^\lambda_{12}$. Such a scenario is however in clear contradiction with $\Delta F=2$ data.

In the case of a significant mass splitting, see left plot in figure \ref{fig:mchi-D11}, we therefore obtain both an upper and a lower bound on the coupling $D_{\lambda,11}$ for small DM masses. The case of quasi degenerate first and third DM generation is even more interesting, as can be seen from the right plot in figure \ref{fig:mchi-D11}. Again the required destructive interference between box and penguin contributions constrains $D_{\lambda,11}$ to a certain range. At the same time the near-degeneracy in mass, due to the DMFV expansion \eqref{eq:mchi} is related to a near-degeneracy $D_{\lambda,11}\simeq D_{\lambda,33}$. With the relic abundance constraint requiring these couplings to be large, DM masses below  $m_{\chi_b}\simeq 100\gev$ are ruled out by the interplay of the various constraints in this scenario.

\begin{figure}[h!]
\centering
\begin{minipage}{6.2cm}
\includegraphics[width=\textwidth]{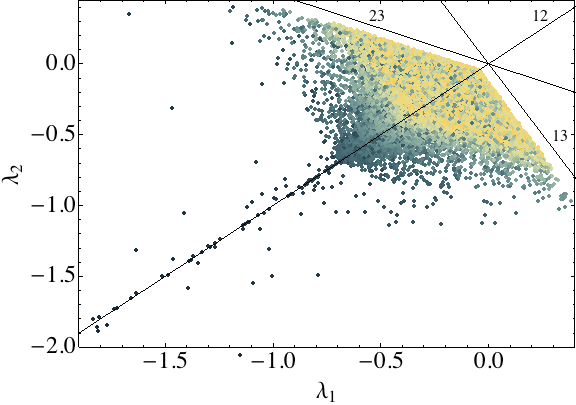}
\end{minipage}\hspace{1mm}
\begin{minipage}{6.2cm}
\includegraphics[width=\textwidth]{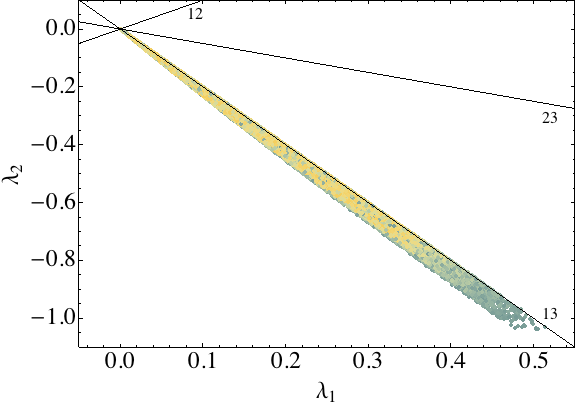}
\end{minipage}\hspace{1mm}
\begin{minipage}{1.8cm}
\includegraphics[width=\textwidth]{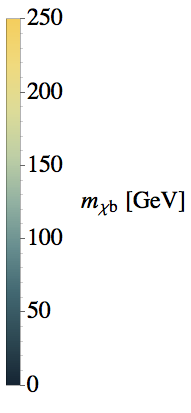}
\vspace*{.01cm}

\end{minipage}
\caption{\label{fig:lambda1-lambda2}Allowed region in the
$\lambda_1$-$\lambda_2$-plane for significant mass splitting $m_{\chi_{d,s}}>1.1m_{\chi_b}$ scenario (left), and in the $13$-degeneracy scenario $m_{\chi_{d}} \simeq m_{\chi_b}$ (right), after imposing the relic abundance, LUX and flavor
constraints. The DM mass $m_{\chi_b}$ is indicated by the color, and the $ij$-degeneracy lines are sketched.}
\end{figure}

Figure \ref{fig:lambda1-lambda2} visualises the allowed non-degeneracies in the coupling matrix $D_\lambda$, compare the parameterisation in \eqref{eq:Dlambda}.
We see that the two scenarios considered correspond to different regions for the parameters $\lambda_{1,2}$. We also observe that with decreasing DM mass the allowed region moves away further from the universality point $\lambda_1=\lambda_2=0$.

\section{LHC signatures}

With the WIMP miracle in mind, the new particles introduced by the mDMFV model are expected to be around the electroweak scale and therefore accessible to the LHC. On the one hand we have the three generations of $\chi_i$ fermions, and on the other hand the scalar mediator $\phi$ which is coloured and will therefore be copiously produced. 

Due to the unbroken $\mathbb{Z}_3$ symmetry these particles have to be pair produced, and the lightest state (assumed to be $\chi_b$) escapes detection and leads to missing energy signatures. The heavier $\chi_i$ states, due to the small mass splitting, give rise to soft jets or photons when decaying into the lightest state. Since these are usually not considered in experimental analyses, we can treat the three $\chi_i$ states equally as missing energy states. The scalar mediator $\phi$ on the other hand decays into a quark jet and missing energy, with the composition of quark flavours given by the size of the couplings $D_{\lambda,ii}$.

\begin{figure}[!h]
  \begin{center}
    \includegraphics[width=0.45\textwidth]{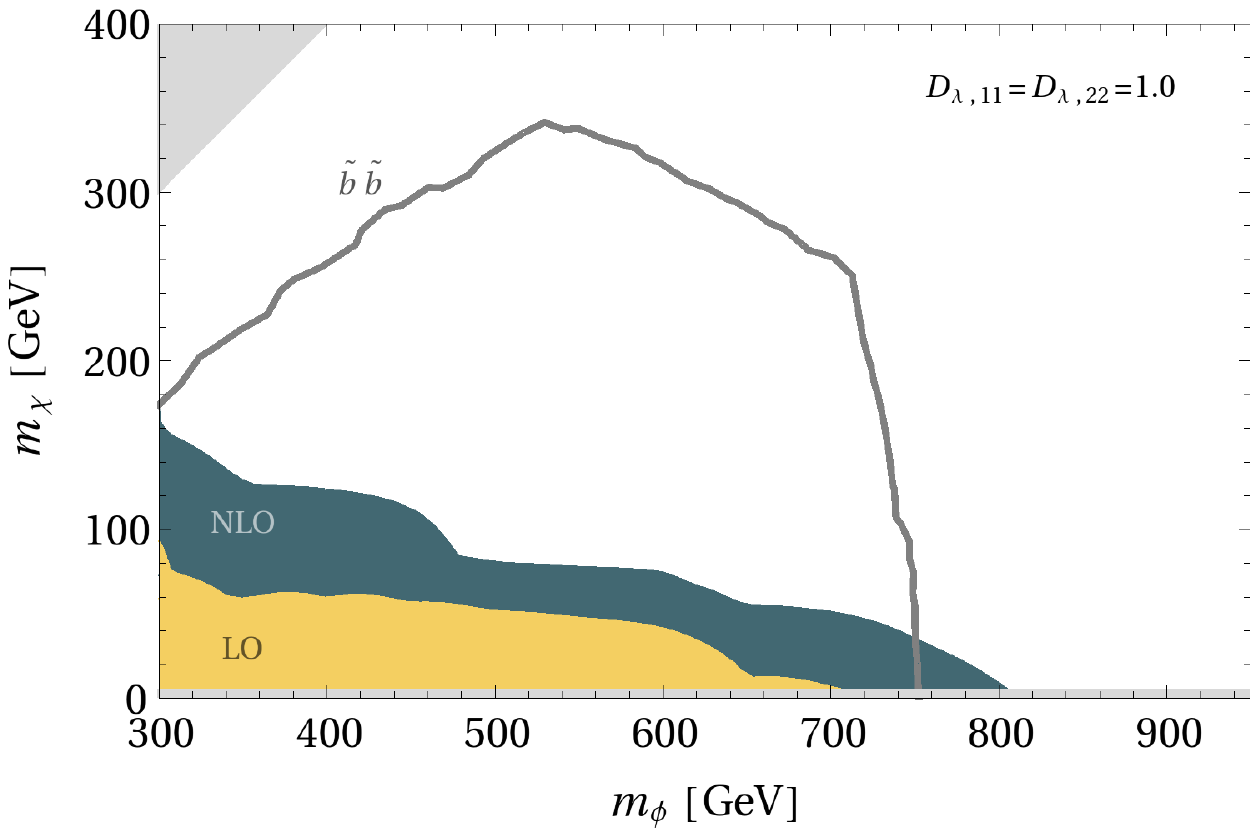}
    \quad
    \includegraphics[width=0.45\textwidth]{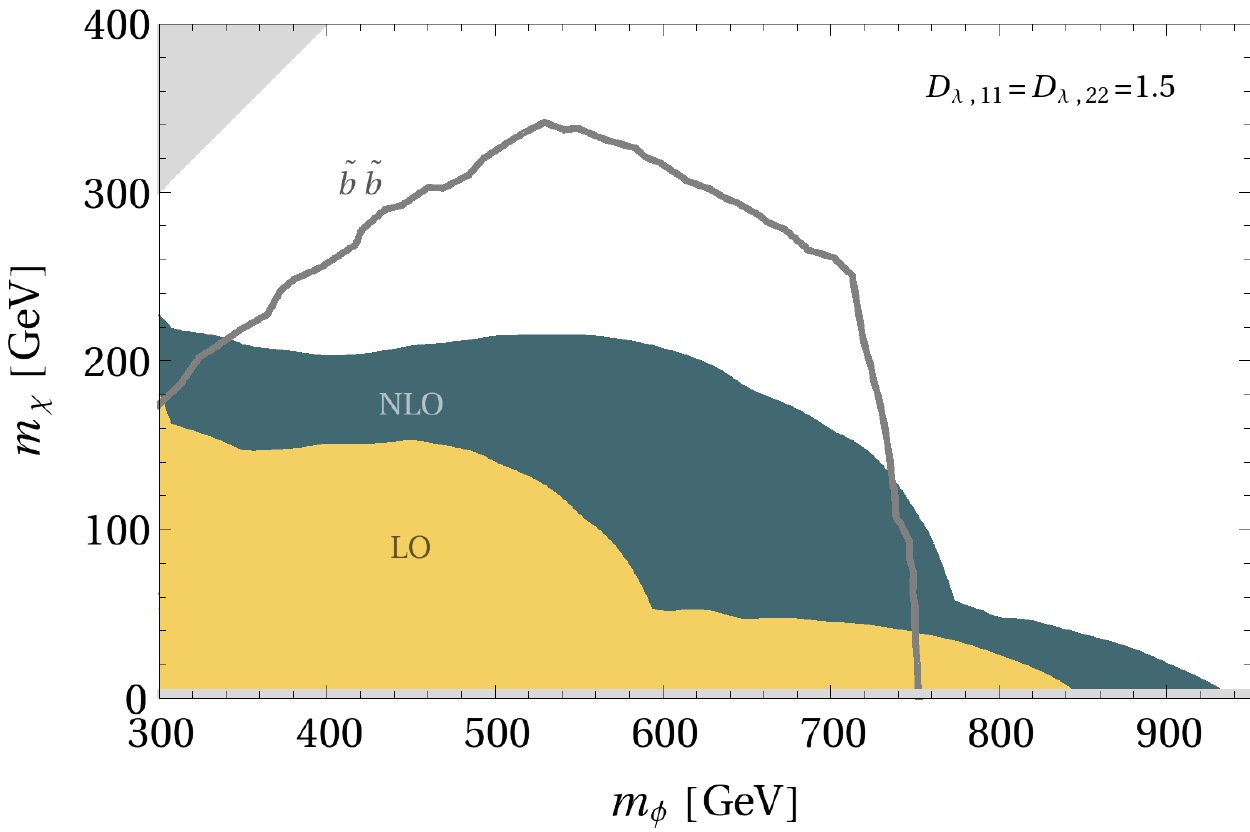}
  \end{center}
  \caption{Limits on the mDMFV model from the CMS 19.5
  fb$^{-1}$ sbottom~\cite{CMS:2014nia} and
  squark~\cite{Chatrchyan:2014lfa} search.  The DM coupling to $b$-quark, $D_{\lambda,33}$
  is fixed everywhere by the corresponding relic abundance
  constraint, and all mixing angles are set to zero for simplicity.}
  \label{fig:sbottoms1}
\end{figure}

The resulting DMFV signatures are thus very similar to the ones from popular SUSY models with $R$-parity conservation. In particular $\phi$ pair production is constrained by sbottom and light squark searches, due to the analogous final state signature. Adapting the  CMS 19.5
  fb$^{-1}$ sbottom~\cite{CMS:2014nia} and
  squark~\cite{Chatrchyan:2014lfa} search to the mDMFV model by taking into account both the changes in production cross section and branching ratios, we find the estimated exclusion contours shown in figure \ref{fig:sbottoms1}.
For light DM masses these bounds reach up to mediator masses of $m_\phi \simeq 850\gev$, however they become significantly weaker for increasing $m_\chi$ mass.

\begin{figure}[!h]
  \begin{center}
    \includegraphics[width=0.45\textwidth]{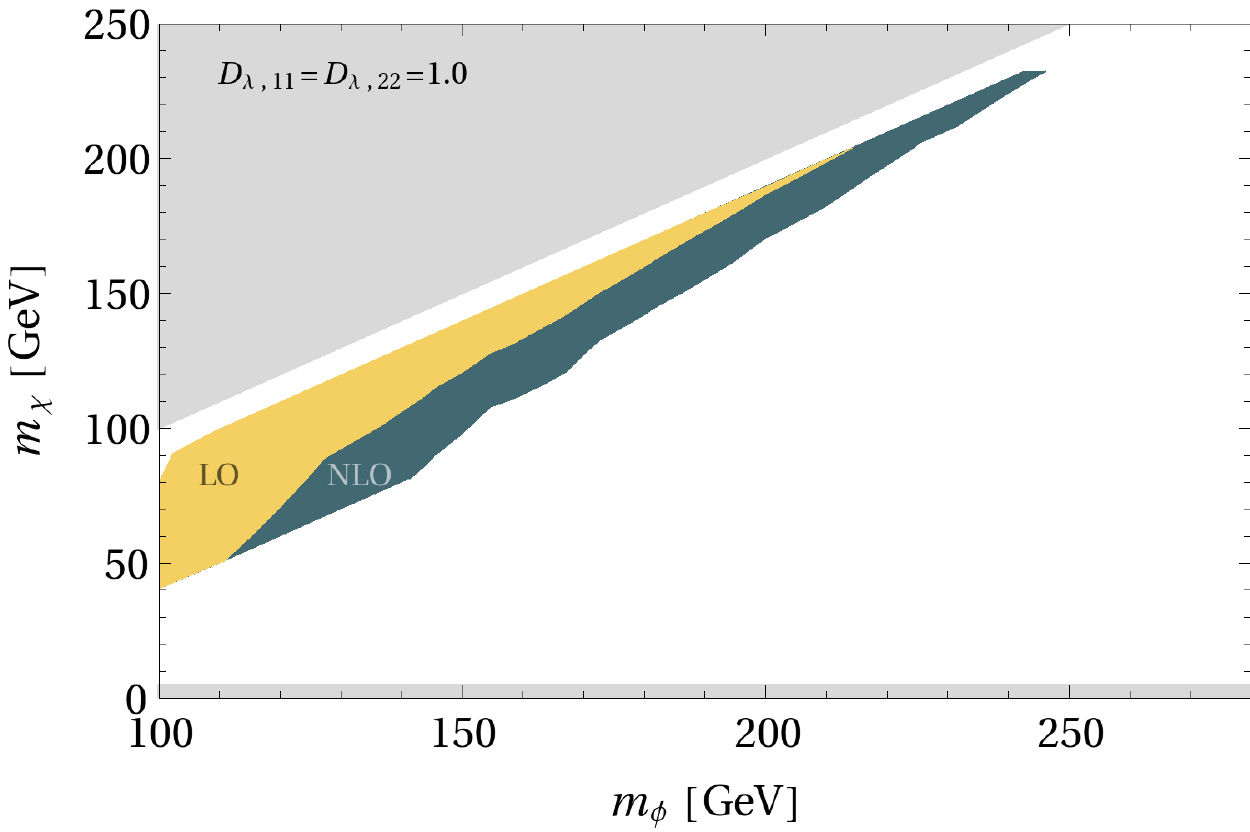}
    \includegraphics[width=0.45\textwidth]{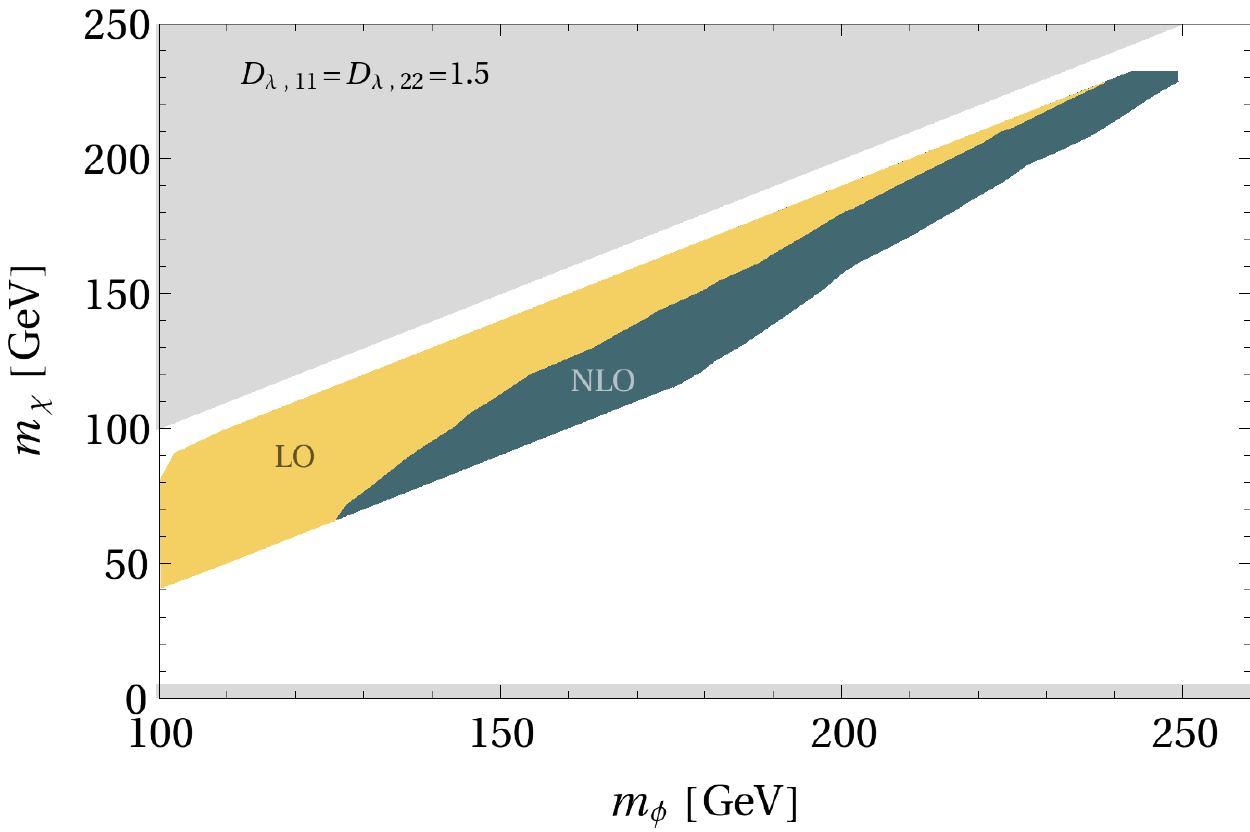}
  \end{center}
  \caption{Limits on the mDMFV model from the CMS 19.7
  fb$^{-1}$ search~\cite{CMS:2014yma} for stops decaying to a
  charm and a neutralino,
  using the monojet + MET final state. 
  }
  \label{fig:monojets}
\end{figure}

Complementary constraints can be obtained from recasting monojet plus missing energy (MET) searches. If the splitting between $m_\phi$ and $m_\chi$ is small, then the $\phi$ decay products are too soft to be observed. This case can be probed in the monojet channel, with the single hard jet arising from initial state radiation. Matching the mDMFV production cross section to the SUSY scenario studied in \cite{CMS:2014yma} yields the constraints displayed in figure \ref{fig:monojets}.

\begin{figure}[!h]
  \begin{center}
    \includegraphics[width=0.45\textwidth]{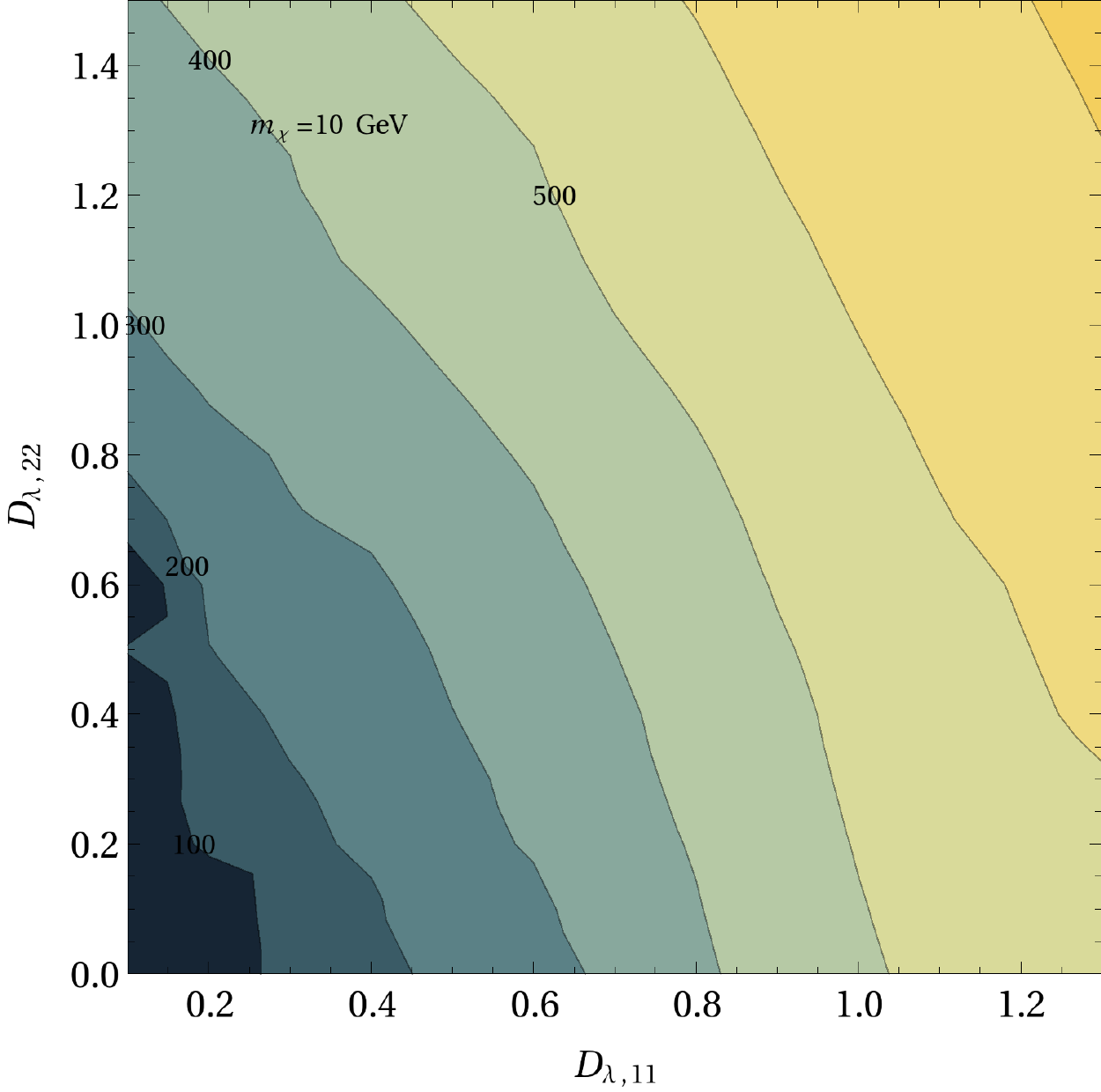}
    \qquad
    \includegraphics[width=0.45\textwidth]{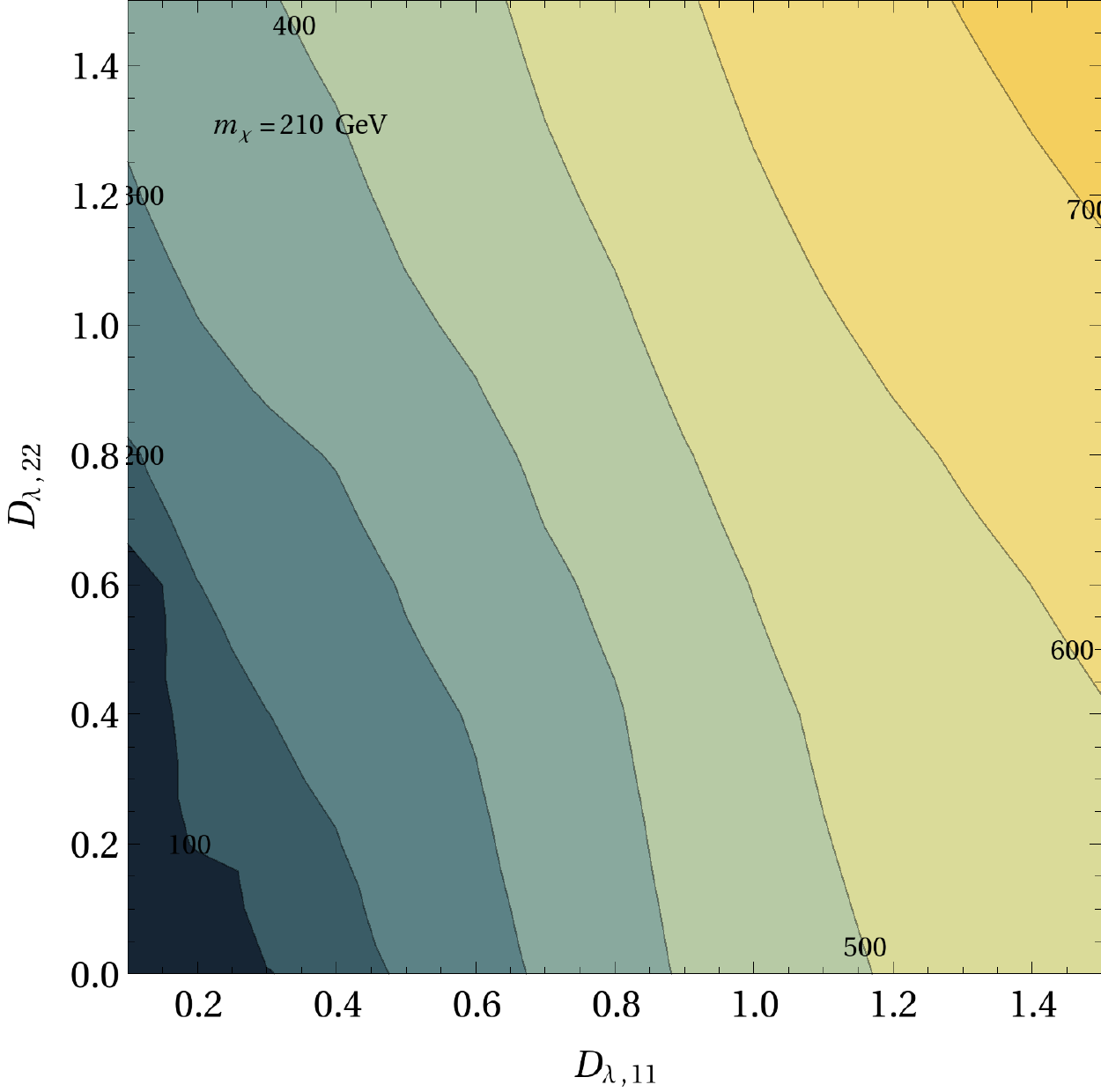}
  \end{center}
  \caption{Limits on the mDMFV model in terms of mass of $\phi$ (in
  GeV) from the CMS monojet
  search~\cite{CMS:rwa}. }
  \label{fig:monojetscoup}
\end{figure}

Monojet searches are also sensitive to direct DM pair production, as well as to pair production of the heavier $\chi$ flavours. Adapting the results of \cite{CMS:rwa} to the production cross section predicted in the mDMFV model, we obtain the constraints on the mediator mass $m_\phi$ shown in figure \ref{fig:monojetscoup}. In contrast to the dijet constraints, in this case the obtained result is rather insensitive to the value of the DM mass $m_\chi$.

In future searches, apart from the channels discussed above, the mDMFV model gives rise to some interesting signatures. Similarly to the case of flavour violating squark decays \cite{Blanke:2013uia}, $\phi$ pair production and decay can lead to one $b$-jet and one light jet in the final state, accompanied by missing energy. With the stringent flavour physics constraint, such a signature is unlikely to be generated by flavour violating bottom squark coupling.
Additionally the soft decay products of the heavier DM flavour would be interesting to look for, in particular since due to the small mass splitting the decay length is sizeable and could be observed as a displaced decay vertex.

\section{Summary}

In analogy to SM matter it is theoretically well-motivated to consider that also the DM carries flavour quantum number and comes in multiple copies. Phenomenologically this scenario becomes particularly rich when non-minimally flavour violating interactions with the SM quarks are considered. The data on FCNC processes provides interesting constraints on the structure of the DM coupling matrix, while the mass spectrum can be constrained mostly from direct DM and LHC searches. A study of the interplay of the various sectors yields the most useful information on the model parameters.


\begin{thebibliography}{99}




%\cite{Kile:2013ola}
\bibitem{Kile:2013ola}
  J.~Kile,
  %``Flavored Dark Matter: A Review,''
  Mod.\ Phys.\ Lett.\ A {\bf 28} (2013) 1330031
  [arXiv:1308.0584 [hep-ph]].
  %%CITATION = ARXIV:1308.0584;%%
  %6 citations counted in INSPIRE as of 25 Nov 2014



%\cite{Buras:2000dm}
\bibitem{Buras:2000dm}
  A.~J.~Buras, P.~Gambino, M.~Gorbahn, S.~Jager and L.~Silvestrini,
  %``Universal unitarity triangle and physics beyond the standard model,''
  Phys.\ Lett.\ B {\bf 500} (2001) 161
  [hep-ph/0007085].
  %%CITATION = HEP-PH/0007085;%%
  %410 citations counted in INSPIRE as of 24 Nov 2014

%\cite{D'Ambrosio:2002ex}
\bibitem{D'Ambrosio:2002ex}
  G.~D'Ambrosio, G.~F.~Giudice, G.~Isidori and A.~Strumia,
  %``Minimal flavor violation: An Effective field theory approach,''
  Nucl.\ Phys.\ B {\bf 645} (2002) 155
  [hep-ph/0207036].
  %%CITATION = HEP-PH/0207036;%%
  %917 citations counted in INSPIRE as of 24 Nov 2014


%\cite{Agrawal:2014aoa}
\bibitem{Agrawal:2014aoa}
  P.~Agrawal, M.~Blanke and K.~Gemmler,
  %``Flavored dark matter beyond Minimal Flavor Violation,''
  JHEP {\bf 1410} (2014) 72
  [arXiv:1405.6709 [hep-ph]].
  %%CITATION = ARXIV:1405.6709;%%
  %6 citations counted in INSPIRE as of 21 Nov 2014

%\cite{Batell:2011tc}
\bibitem{Batell:2011tc}
  B.~Batell, J.~Pradler and M.~Spannowsky,
  %``Dark Matter from Minimal Flavor Violation,''
  JHEP {\bf 1108} (2011) 038
  [arXiv:1105.1781 [hep-ph]].
  %%CITATION = ARXIV:1105.1781;%%
  %29 citations counted in INSPIRE as of 28 Nov 2014


%\cite{Blanke:2006xr}
\bibitem{Blanke:2006xr}
  M.~Blanke, A.~J.~Buras, A.~Poschenrieder, S.~Recksiegel, C.~Tarantino, S.~Uhlig and A.~Weiler,
  %``Another look at the flavour structure of the littlest Higgs model with T-parity,''
  Phys.\ Lett.\ B {\bf 646} (2007) 253
  [hep-ph/0609284].
  %%CITATION = HEP-PH/0609284;%%
  %75 citations counted in INSPIRE as of 10 Dec 2014


%\cite{Agrawal:2014una}
\bibitem{Agrawal:2014una}
  P.~Agrawal, B.~Batell, D.~Hooper and T.~Lin,
  %``Flavored Dark Matter and the Galactic Center Gamma-Ray Excess,''
  Phys.\ Rev.\ D {\bf 90} (2014) 063512
  [arXiv:1404.1373 [hep-ph]].
  %%CITATION = ARXIV:1404.1373;%%
  %38 citations counted in INSPIRE as of 09 Dec 2014

%\cite{Akerib:2013tjd}
\bibitem{Akerib:2013tjd}
  D.~S.~Akerib {\it et al.}  [LUX Collaboration],
  %``First results from the LUX dark matter experiment at the Sanford Underground Research Facility,''
  Phys.\ Rev.\ Lett.\  {\bf 112} (2014) 9,  091303
  [arXiv:1310.8214 [astro-ph.CO]].
  %%CITATION = ARXIV:1310.8214;%%
  %503 citations counted in INSPIRE as of 09 Dec 2014


%\cite{CMS:2014nia}
\bibitem{CMS:2014nia}
  CMS Collaboration [CMS Collaboration],
  %``Search for direct production of bottom squark pairs,''
  CMS-PAS-SUS-13-018.
  %%CITATION = CMS-PAS-SUS-13-018;%%
  %18 citations counted in INSPIRE as of 08 Dec 2014


%\cite{Chatrchyan:2014lfa}
\bibitem{Chatrchyan:2014lfa}
  S.~Chatrchyan {\it et al.}  [CMS Collaboration],
  %``Search for new physics in the multijet and missing transverse momentum final state in proton-proton collisions at $\sqrt{s}$= 8 TeV,''
  JHEP {\bf 1406} (2014) 055
  [arXiv:1402.4770 [hep-ex]].
  %%CITATION = ARXIV:1402.4770;%%
  %71 citations counted in INSPIRE as of 08 Dec 2014

%\cite{CMS:2014yma}
\bibitem{CMS:2014yma}
  CMS Collaboration [CMS Collaboration],
  %``Search for top squarks decaying to a charm quark and a neutralino in events with a jet and missing transverse momentum,''
  CMS-PAS-SUS-13-009.
  %%CITATION = CMS-PAS-SUS-13-009;%%
  %17 citations counted in INSPIRE as of 08 Dec 2014



%\cite{CMS:rwa}
\bibitem{CMS:rwa}
  [CMS Collaboration],
  %``Search for new physics in monojet events in pp collisions at sqrt(s)= 8 TeV,''
  CMS-PAS-EXO-12-048.
  %%CITATION = CMS-PAS-EXO-12-048;%%
  %103 citations counted in INSPIRE as of 08 Dec 2014

%\cite{Blanke:2013uia}
\bibitem{Blanke:2013uia}
  M.~Blanke, G.~F.~Giudice, P.~Paradisi, G.~Perez and J.~Zupan,
  %``Flavoured Naturalness,''
  JHEP {\bf 1306} (2013) 022
  [arXiv:1302.7232 [hep-ph]].
  %%CITATION = ARXIV:1302.7232;%%
  %31 citations counted in INSPIRE as of 10 Dec 2014


\end{thebibliography}
\end{document}